\documentclass[final,1p,times]{elsarticle}

 \biboptions{comma,square}

\usepackage[T1]{fontenc}
\usepackage[latin9]{inputenc}
\usepackage{verbatim}
\usepackage{amstext}
\usepackage[usenames]{color}
\usepackage{graphicx,psfrag}
 \usepackage{amssymb}

\def\ie{\emph{i.e. }}

\def\beq{\begin{equation}}
\def\eeq{\end{equation}}
\def\beqa{\begin{eqnarray}} 
\def\eeqa{\end{eqnarray}}
\def\beg{\begin{lyxgreyedout}}
\def\eeg{\end{lyxgreyedout}}

\def\eps{{\epsilon}}

\journal{Physics Letters A}

\begin{document}

\begin{frontmatter}

\title{The role of quantum recurrence in superconductivity, carbon nanotubes and related gauge symmetry breaking}
  \author[unicam]{Donatello Dolce}
\ead{donatello.dolce@unicam.it}

  \author[unicam]{Andrea Perali}
  
\address[unicam]{University of Camerino, Piazza Cavour 19F, 62032 Camerino, Italy.}

\begin{abstract}
Pure quantum phenomena are characterized by intrinsic recurrences in space and time.  We use such an intrinsic periodicity as a quantization condition to derive the essential phenomenology of superconductivity. The resulting description is based on fundamental quantum dynamics and geometrical considerations, rather than on microscopical characteristics of the superconducting materials. This allows for the interpretation of the related gauge symmetry breaking by means  of the competition between quantum recurrence and thermal noise. We also test the validity of this approach to describe the case of carbon nanotubes. 
\end{abstract}

 \end{frontmatter}

\section*{Introduction}
 In this paper we derive fundamental aspects of SuperConductivity (SC) in terms of simple considerations about the intrinsic recurrence  of quantum systems (also known  in condensed matter as ``complete'' or ``macroscopic'' coherence). 
It is an empirical fact that pure quantum phenomena (such as relativistic isolated elementary particles in which the quantum recurrence is associated to the so-called de Broglie or Compton internal clock) are characterized by recurrences in time and space.  A description of elementary particles as ``periodic phenomena'' was originally suggested by de Broglie  \cite{Broglie:1924}. For instance, this implies that every particle can be regarded as a reference clock  whose rest ticks, of Compton time duration $T_{\tau}$, are determined by the particle's mass $\bar m = h / T_{\tau} c^{2}$, as indirectly confirmed by recent experiments \cite{Lan01022013,2008FoPh...38..659C}. Here  we will mainly consider the temporal component of the recurrence, being particularly convenient for the description of macroscopic quantum phenomena.

In recent publications \cite{Dolce:cycles,Dolce:2009ce,Dolce:tune} we have proven that  fundamental aspects of Quantum Mechanics (QM) can be obtained semi-classically by imposing as constraint the natural recurrences of the elementary particles prescribed by undulatory mechanics (wave-particle duality). By imposing intrinsic periodicity as a constrain by means of Periodic Boundary Conditions (PBCs)  the particle turns out to be as a ``particle in a box''  or  a vibrating string: the harmonics associated to these recurrences can be regarded as the quantum excitations of the system.  The ideas can be regarded as a relativistic generalization  of the Bohr-Sommerfeld quantization or of Bloch's theorem. In Bohr's semi-classically description, the atomic orbitals are given by the  possible vibrational modes with integer number of the of modulated space-time quantum recurrences (closed orbits) of the electron wave-function associated to a Coulomb potential, combined with the spherical harmonics associated to the spherical periodicity \cite{refId0} --- directly observed in a recent experiment \cite{PhysRevLett.110.213001}.  In this paper we will simply apply this quantization prescription to study SC; considerations about the novelty, consistency and applicability of this approach go beyond the scope of this paper and are described in detail in the published papers \cite{Dolce:cycles,Dolce:2009ce,Dolce:tune,Dolce:AdSCFT}.
Similarly, we will see that the constraint of coherent intrinsic periodicity for the Electric Charge Carriers (ECCs) in a conductor ring can be used to derive the most striking quantum aspects of SC. Phenomenological manifestation of the quantum recurrences in SC are described for instance in \cite{PhysRevLett.9.9,Lurie1970,Loder08magneticflux}.  Intuitively, as in Bohr's atom, closed orbits of the wave-function along the circuit implicitly means stability of the corresponding electric current.
 In particular, we will describe the collective wave-function of the ECCs as a sum of the standing waves (\emph{i.e.} harmonic modes) satisfying the PBCs allowed by the symmetry of the system. As a direct consequence of this constraint of intrinsic periodicity for gauge invariant matter fields,  the corresponding Goldstone field, being a phase of a ``periodic phenomenon'', can only vary by discrete amounts during these electron closed orbits along the superconductor. 
 Hence the magnetic flux turns out to be quantized and the current cannot decay smoothly, as also argued from more general geometrodynamical considerations in \cite{Dolce:tune}.  This will allows us to obtain the fundamental aspects of Weinberg's description of SC and thus to describe the exclusion of the magnetic field and the alternating currents at a junction of two superconductors, \emph{i.e.} the Meissner and Josephson effects, in terms of intrinsic periodicity of the ECC wave-function in the superconductors. As known from Bloch's theorem, the intrinsic periodicity of the lattice (and of the related electronic orbitals) also determines the energy bands of the materials. The fundamental energy scale is fixed through the Planck constant by the inverse of the temporal period of the intrinsic recurrence. The effect of  an external magnetic flux is to twist the PBCs, obtaining the Aharonov-Bohm effect and the band gap opening.

Carbon Nanotubes (CNs) (as well as graphene bi-layers and metallic stripes) turn out to be  particularly fortunate physical systems to illustrate and to test the validity of our description. In fact, the quantum recurrence in CNs is directly related to the curled up dimension of the graphene lattice. In particular, starting from the massless dispersion relation of the Dirac cones in pure 2D  graphene layers, the residual temporal periodicity along the curled up dimension for ECCs at rest with respect to the axial direction determines the Compton time of the system and generates the  ECCs effective mass.   

The  {  approach discussed and adopted in} this paper is well described by Weinberg's words \cite{Weinberg:1996kr}: \emph{``A superconductor is simply a material in which ElectroMagnetic (EM) gauge is spontaneously broken. 
[...]. Detailed dynamical theories are needed to explain why and what temperature this symmetry breaking occurs, but they are not needed to derive the most striking aspects of SC: exclusion of magnetic fields, flux quantization, zero resistivity, and alternating currents at a gap between superconductors held at different voltage. 
As we shall see here, these consequences of broken gauge invariance can be worked out [...] solely on the basis of general properties of Goldstone mode''}. We will conclude the paper by proposing a possible interpretation of the EM gauge symmetry breaking  in SC in terms of intrinsic recurrence. As suggested by Weinberg, the discrete values of the Goldstone field denotes a breaking of the EM gauge symmetry --- typically from $U_{EM}(1)$ to $\mathcal Z_{2}$. {  This can happen only at a  sufficiently low temperature. In fact the quantum recurrence, \ie a ``complete coherence'' of the periodicity of the ECC wave-function, is effectively destroyed by the thermal noise, so that the gauge invariance is restored. Thus, to have correlation and in turn symmetry breaking, the duration of the quantum recurrence must be shorter than the characteristic thermal scales.  In our study of SC in terms of intrinsic recurrence, this describes the essential feature of the transition between the coherent state (quantum recurrence) and the normal state (uncorrelated wave-functions)}.  

      It is interesting to note that the description of the quantum behaviour of intrinsically periodic phenomena \cite{Dolce:cycles,Dolce:2009ce,Dolce:tune} used in this paper is similar to that of 't Hooft's model of continuous periodic Cogwheels \cite{hooft:2007determinism,hooft:2014cogwheel}. Similarly CNs represents a realisation of discrete periodic Cogwheels (i.e. a "particle moving very fast of a circle" which is dual to "the quantum harmonic oscillator") in which the number of Cellular Automata cites is the number of carbon atoms in a CN diameter. Indeed the Elementary Cycles theory proposed in \cite{Dolce:cycles,Dolce:2009ce,Dolce:tune} is an effective description of 't Hooft model. Thus the description of the quantum behaviour of SC superconductivity and graphene physics proposed in this paper can be extended to 't Hooft model. In particular  our results is a successful test of these foundational  aspects of quantum mechanics. Due to the periodicity in time of the quantum recurrence, our approach shares also fundamental analogies with the description of SC based on time crystals and related phenomenology \cite{Wilczek:2012crys,Yoshii:2014:suprcry}.   

 The study of SC is also of interest because it is at the origin of the Higgs mechanism \cite{PhysRev.117.648,Goldstone1961,PhysRevLett.13.508,PhysRevLett.13.585} and thus of the generation of particles' mass.  The analogies between the CNs physics and eXtra Dimensional (XD) physics, pointed out in  \cite{deWoul:2012ed} and which has an intuitive interpretation in terms of ``virtual XD'' \cite{Dolce:AdSCFT}, provide interesting correspondences between the gauge symmetry breaking of SC and that of the most investigated extensions of the Higgs mechanism (composite-Higgs, gauge-Higgs unification, Technicolor, and related models) \cite{Sundrum:2005jf,Csaki:2005vy,Arkani-Hamed:2001ca,Son:2003et}. 
Similarly to  the Scherk-Schwartz mechanism \cite{Scherk197961} and Hosotani \cite{Hosotani1983309}  the symmetry breaking can be interpreted as induced by the BCs, which in our case encode the quantum recurrence. 

\section{Quantum recurrence}\label{EM:time:period}
As described in recent papers \cite{Dolce:cycles,Dolce:2009ce,Dolce:tune}, and in analogy with Bloch's theorem, our ansatz to investigate the quantum phenomenon of SC is to assume that the ECCs, \ie the electrons responsible of conductivity in ordinary materials, are characterized by  {  collective intrinsic temporal recurrences. For simplicity  we consider the case of a single generic temporal recurrence} $T(\bar p_{\parallel})$ determined by the microscopic (crystalline) properties of the material {--- more periodicities at different scales should be considered for more specific phenomenological descriptions}. For instance, this fixes the shape of energy bands, as we will also see for  CNs. It depends on the momentum   $\bar p_{\parallel} = h / \lambda_{\parallel}$ of the ECCs in the direction of the current,  \ie by the spatial recurrence $\lambda_{\parallel}$ along the axial direction of the conductor. Both the spatial and temporal periodicities are derivable from the proper time periodicity (Compton time) $T_{\tau} = T(0)$ of the ECCs in the material, typically representing the extremal limit of the temporal recurrence. 
For simplicity's sake we start by only considering the temporal recurrence as this is sufficient to characterize the novelty of the approach presented here, but analogous conclusions can be achieved in terms of spatial recurrence through the equations of motion. Our description is therefore analogous to 't Hooft continuous periodic Cogwheel model. 

Thus we assume that a generic collective wave-function of the ECCs in the superconducting material satisfies the following PBCs 
\beq 
\psi(\mathbf{x},t) = \psi(\mathbf{x},t + T(\bar  p_{\parallel})).\label{Period:BCs}
\eeq 
Beside this we can assume, for instance, discrete symmetries, such as a symmetry for inversion $t + T/2 \leftrightarrow -t + T/2$, in analogy  a vibrating string {  of length $L$ in which the possible vibrational wave-numbers are $k_{n} = n \bar k = n / 2 L$. This means that the periodicity of the phase is $n \pi$ rather than $2 \pi n$ and it} can be represented by a Orbifold $\mathbb S^{1}/\mathcal Z_{2}$.  
 In this case, upon EM gauge invariance $U_{EM}(1)$ of the ECC 
\begin{equation}
\psi'(\mathbf{x},t) = U(\mathbf{x},t)\psi(\mathbf{x},t)  \,, \,\,  \text{where} \, \, U(\mathbf{x},t)=e^{-i\frac{e}{\hbar c}\theta(\mathbf{x},t)}~,\label{gauge:inv:scond} \label{gauge:transf:scond}
\end{equation}
 the intrinsic recurrence implies that the phase of the gauge transformation has an induced periodicity and is defined modulo factors  $ \pi n $: 
\begin{equation}
\frac{e}{c \hbar}\theta(\mathbf{x},t)=\frac{e}{c \hbar}\theta(\mathbf{x},t + T(\bar p_{\parallel}))+ \pi n ~.\label{eq:periodic:goldstone:phase}
\end{equation} 
As a consequence of this recurrence the Goldstone field of the gauge transformation can vary only by finite amounts  
\begin{equation}
{\Delta {\theta}}(\mathbf{x},t) = \frac{\phi_{0} }{2 } \,, \,\,  \text{where} \, \phi_{0} = \frac{h c} {e}\,.\label{goldstone:periodic}
\end{equation}   
This shows that the discrete variations of the Goldstone field of Weinberg's description of SC \cite{Weinberg:1996kr} has indeed a simple physical justification in terms of intrinsic periodicity.

\subsection{Flux quantization}
 By following the line of \cite{Weinberg:1996kr}, in a superconductor with the shape of a torus we can draw a contour $\Sigma$ on which the EM field is a pure gauge 
\beq
A_{\mu}(\mathbf{x},t)=\partial_{\mu}{{\theta}}(\mathbf{x},t)\,.\label{pure:tgauge}
\eeq
According to our ansatz (\ref{goldstone:periodic}), for closed orbits of the wave-function,  ${\theta}(\mathbf{x},t)$ can only vary
by discrete amounts running along the contour $\Sigma$. In this way we find that the Stokes theorem gives a quantization
of the magnetic flux through the area $\mathcal{S}_{\Sigma}$ limited by $\Sigma$,  
\begin{eqnarray}
 \int_{\mathcal{S}_{\Sigma}}\!\! \mathbf{B}(\mathbf{x},t)\cdot d\mathbf{S} 
= \oint_{\Sigma_{\mathcal{}}}\!\! \mathbf{A}(\mathbf{x},t)\cdot d\mathbf{x} 
  =  \oint_{\Sigma}\!\! \mathbf{\nabla}{{\theta}}(\mathbf{x},t)\cdot d\mathbf{x}=n \frac{   \phi_{0}}{2 } 
  \,. 
  \label{magn:quant}
  \end{eqnarray}
  
  We obtain therefore that the magnetic flux in a superconducting ring is quantized, and the quantum unit is $h c / 2 e$, as well known in SC. 
 Since the magnetic flux is quantized, the electric current cannot
smoothly decay while flowing around the torus, and there is not electric resistance. This peculiar behavior is indeed characteristic of SC. This is manifestly  a quantum effect since it disappears in the classical limit $\hbar \rightarrow 0$. Such a quantization of the magnetic flux has been obtained without any further quantization prescription than intrinsic periodicity \cite{Dolce:cycles,Dolce:2009ce,Dolce:tune}. The number $n$ of nodes of the wave function determines the number of vortices that can be generated by the supercurrent.

As noted in \cite{Weinberg:1996kr}, the behavior of the Goldstone field $\theta(\mathbf x, t)$ suggests the link between our description of SC and the conventional Ginzburg-Landau formulation.  In fact, the quantization of $\theta(\mathbf x, t)$  means that 
the Goldstone field transforms as the phase of a condensate of a fermionic pair operator $\left\langle\eps_{\alpha \beta} \psi^{\alpha} \psi^{\beta}\right\rangle$  of charge $-2e$,  according to (\ref{goldstone:periodic}).  That is, it plays the role of the usual Cooper pair of the BCS microscopic theory of SC. As we will discuss also in sec.(\ref{Gauge:symmetry:breaking}), at sufficiently low temperature such a quantization of the Goldstone field breaks the $U_{EM}(1)$ gauge invariance to $\mathcal Z_{2}$. 

We can more generally assume a symmetry by rotations of an angle $2 \pi \alpha$, where $\alpha \in \mathbb R$. This corresponds to twist the PBCs
\beq 
\psi(\mathbf{x},t) = e^{ i 2 \pi \alpha}\psi(\mathbf{x},t + T(\bar  p_{\parallel}))\,.\label{twisted:PBCs}
\eeq 
In this case, assuming periodicity $\mathbb S^{1}$ (we are now considering a phase periodicity is $2 \pi n$ and not $\pi n$ as in $\mathbb S^{1}/\mathcal Z_{2}$), the resulting quantization of the magnetic flux is 
\beq 
\int_{\mathcal{S}_{\Sigma}}\mathbf{B}(\mathbf{x},t)\cdot d\mathbf{S}
=   
\left( n +   \alpha \right) \phi_{0}   
~.\label{magn:quant}
\eeq 
This case will correspond to a band gap opening. The quantized magnetic flux is shifted and the quantized levels of the Goldstone field has a non vanishing ground value 
\beq \bar{\theta} (\mathbf x, t)=  \alpha  \phi_{0}  
\,.
\eeq
The PBCs in (\ref{Period:BCs}) corresponds to $\alpha = 0$ whereas for anti-PBCs (reflecting the spin-statistic of fermionic particles) we have $\alpha = \frac{1}{2}$. We assume $n \in \mathbb N$, that is a fixed direction of the current and thus of the magnetic field. If $\alpha =0 $ we can neglect the mode $n=0$ as it corresponds to the configuration with zero magnetic flux in the superconductor, according to (\ref{magn:quant}). In fact, once that an external magnetic field has induced a current  ($n\geq 1$), it cannot decay to the mode with $n=0$ for symmetry reasons\footnote{The semiclassical quantization of the magnetic flux induced by the PBCs is similar to the quantization of the angular momentum in a cylindric Schr\"odinger problem. The quantum number $n$ can also be regarded as describing the quantized angular momentum of the current. Thus, for momentum conservation, once that a current is created $n \geq 1$ it cannot decay to the state $n=0$. A similar effect can be observed in superfluidity of a toroidal Bose-Einstein condensate where the case with zero angular momentum is only possible by introducing a potential barrier that breaks the rotational invariance of the superflow \cite{2013JPhB...46i5302P}.}.

As we will see,  the description of the magnetic flux quantization can also be directly tested in CNs. 

\subsection{{Meissner and Josephson effect}}\label{Meissner:Josephson:effect} 
 As can be seen by assuming the temporal gauge,  the energy of a given configuration of the EM field is described by the time derivative term in the EM Lagrangian,  $\mathcal{S}_{EM}$. 
 Under a gauge transformation this energy term transforms as 
 \begin{eqnarray}
 A^{i,0}A_{i,0} \rightarrow  A^{i,0}A_{i,0} -    \partial_{i}{{\theta}} \partial^{i}{{\theta}}   ~.\label{Meissner:tgauge:YM}
\end{eqnarray}

Now we must consider that along the contour $\Sigma$ of the superconductor, the EM field is pure gauge (\ref{pure:tgauge}).  
Beside the quantization of the magnetic flux, we have that the energy of the periodic EM field has a local minimum  deep inside a large superconductor and the
magnetic field vanishes  ($\mathbf{B}\equiv0$), such that  
$F_{\mu\nu}(\partial_{i}{\theta})|_{\Sigma}=0$.
The consequence of (\ref{Meissner:tgauge:YM}) is that, inside the superconductor, the energy cost to expel the magnetic
field is small with respect to the energy of a configuration where
the magnetic field is inside the superconductor. 
That is, \cite{Weinberg:1996kr}, close to the local energy minimum (\ref{pure:tgauge}),
the EM Lagrangian  can be expanded as \beq 
\mathcal{L}_{EM}[A_{\mu}-\partial_{\mu}{\theta}]\sim\mathcal{L}_{EM}[0]+\frac{\partial^{2}\mathcal{L}_{EM}[A_{\mu}-\partial_{\mu}{\theta}]}{2\partial(A_{\mu}-\partial_{\mu}{\theta})^{2}}(A_{\mu}-\partial_{\mu}{\theta})^{2} \,.\label{Meissner:expansion}\eeq 
The leading term of this expansion is zero in the pure gauge configuration on $\Sigma$. The coefficient of
the quadratic term can be written as a function of the length $\Lambda$. Therefore, in the static case \begin{equation}
\mathcal{L}_{EM}(A_{\mu}-\partial_{\mu}{\theta})
\sim\frac{V}{\Lambda^{2}}(A_{\mu}-\partial_{\mu}{\theta})^{2}\,,\label{Meissner:eq}\end{equation}
where $V$ is the volume of the superconductor. The so-called penetration length $\Lambda$ describes
 the region of the superconductor over which the magnetic field is non zero and is a characteristic of the superconducting material. As we will see, $\Lambda$ is related to the mass of the photon in the supercundoctor.

To describe the Josephson effect we consider a junction between two
superconductors separated by a thin isolating barrier.
By gauge invariance the Lagrangian at the junction can only depend 
on the phase difference of the two Goldstone fields
$ 
\mathcal{L}=\mathcal F(\Delta{\theta}) 
$. 
Since the Goldstone fields is periodic and can vary only by steps $ n  \phi_{0} / 2  $,    $\mathcal F$ is a periodic function as well. In the Orbifold $\mathbb S^{1} / \mathcal Z_{2}$ we have
$
\mathcal F(\Delta{\theta})= \mathcal F\left(\Delta{\theta}+ n\frac{  \phi_{0}}{2 }  \right )
$. 
It is known that this is sufficient to explain  both
the \emph{DC} and \emph{AC} Josephson effects \cite{Weinberg:1996kr}.  At zero voltage difference across the two superconductors 
continuous current flows, depending on the phase difference between
the two Goldstone fields. If a constant voltage difference  $\Delta V$
is maintained between the two superconductors, it implies that an alternate
current flows with fundamental frequency $ \bar f_{junct.} = \frac{1}{T_{junct.}}  = \frac{2 e |\Delta V|}{ h} $. 
 In this way both the Meissner and Josephson effects have been inferred  in terms of the quantum recurrence. 
 
\subsection{{Energy bands structure}}\label{Aharonov:Bohm}
For a relativistic description we must consider that the recurrence is in general characterized by a spacetime periodicity $T^{\mu}=\{T, \vec \lambda / c \}$ associated, through the Planck constant, to the four momentum $\bar p_{\mu}= \{\bar E / c, - \bar \mathbf p \}$ of the ECCs, according to the relation ${\bar{p}}_{\mu}  c T^{\mu} = \hbar$, see \cite{Dolce:cycles} for a review. Thus, by considering both the temporal and spatial periodicities, the relativistic PBCs for the ECCs are  
\beq 
\psi(x^{\mu}) =  \psi(x^{\mu} + c T^{\mu}). \label{Period:BCs:4D}
\eeq 
The effective mass of the ECCs is determined by the proper time periodicity, also known as the Compton time, $T_{\tau} = T(0) = h / \bar m c^{2}$. The gauge connection  between two spacetime points is  
\beq 
\psi(x + x' ) = e^{i \frac{e}{c \hbar} \int_{x}^{x'} A_{\mu} d x^{\mu}} \psi(x),\label{gauge:connection}  
\eeq 
so that, in a Orbifold $\mathbb S^{1}/\mathcal Z_{2}$, the PBCs (\ref{Period:BCs:4D}) leads to the Dirac quantization condition for magnetic monopoles  
\beq 
\oint_{\Sigma} A_{\mu} dx^{\mu} = n \frac{\phi_{0}}{2}. \label{dirac:quant} 
\eeq 

The spatial component of this relation  yields the quantized magnetic flux (\ref{magn:quant}), whereas  the temporal component describes the Josephson effect already discussed in sec(\ref{Meissner:Josephson:effect}). In fact,  we can now consider the temporal component of (\ref{dirac:quant}) together with  the pure gauge condition (\ref{pure:tgauge}) for $A_{0}(x)$ in the junction region (in this case the boundaries are given by the walls of the isolating barrier). Through Stokes' theorem, for a Orbifold $\mathbb S^{1}/\mathcal Z_{2}$ and considering that a voltage difference $\Delta V$ can be written as $\partial_{t} \theta(x) = -  \Delta V$,  a stationary  ECC wave-function between the two superconductors of the junction is such that $ T_{junct.} \Delta V / c =  \phi_{0} / 2 $. Thus we find again the fundamental frequency  $\bar f_{junct.} = 1 / T_{junct.}$ of the Josephson effect.
  From (\ref{dirac:quant}) we finally see that an external magnetic field of flux $\alpha' \phi_{0}$ leads to a shift of the quantized magnetic flux  $(\frac n 2 + \alpha') \phi_{0}$ as prescribed by the Aharonov-Bohm effect. The effect of the external magnetic field is therefore equivalent to a twist of an angle $2 \pi \alpha'$ on the PBCs, similarly to (\ref{twisted:PBCs}). The reason is that gauge interactions can be in general described as modulations of spacetime periodicities, as proven in \cite{Dolce:tune} in terms of spacetime geometrodynamics. Such periodic variations of the magnetic flux associated to the intrinsic periodicity of the ECCs lead to corresponding variations of the critical temperature, i.e. to the Little-Parks effect.

The constraint of periodicity  implies that the ECCs wave-function is the superposition of all the possible harmonics allowed by the PBCs as for a vibrating closed string, for the energy bands of Bloch's theorem or for the orbitals in Bohr's atom. In the continuum, assuming a symmetry $\mathbb S^{1}$, by discrete Fourier transform, the intrinsic time periodicity $T(\bar p_{\parallel})$ implies the harmonic quantized energy spectrum \footnote{By considering the relativistic modulation  of time periodicity  (relativistic Doppler effect) associated to boosts of an elementary particle, the resulting dispersion relation of the energy spectrum is $E_{n}(\bar \mathbf p ) =   n \bar E (\bar \mathbf p ) = n h / T(\bar \mathbf p ) = n \sqrt{ \bar m ^{2} c^{4} + {\bar \mathbf p }^{2} c^{2}}$. The energy bands of a material are the analogous of the quantum levels of the energy spectrum of a second quantized field. The vacuum energy  $n h / 2 T(\bar \mathbf p )$ corresponds to half twist, \emph{i.e.} anti-PBCs. Interaction between the ECCs and the atom in the lattice can result in deformations of this perfectly harmonic band structure. }
 \beq 
 E_{n}(\bar  p_{\parallel}) =    n \bar E (\bar  p_{\parallel}) = n \frac{h}{ T(\bar  p_{\parallel})} \,. 
 \eeq
 The modulation of the temporal recurrence $T(\bar  p_{\parallel})$ for ECCs moving with $\bar  p_{\parallel}$ along the superconductors describes the energy bands structure of the material in the continuous lattice  limit. In particular, this case corresponds to metallic materials. 
 
The case of semiconducting materials is given by twisted-PBCs so that the energy spectrum is shifted by a factor $\alpha \bar E $. In fact the quantization condition is $\exp[- i E_{n}(p_{\parallel}) T(p_{\parallel})/ \hbar] = \exp[ -i 2 \pi (n + \alpha)]$. Thus the resulting energy spectrum associated to the temporal recurrence for the ECCs in their motion along the superconductors  has a band gap opening
 \beq E_{n}(p_{\parallel}) = \left(n + \alpha \right) \frac{h}{T(p_{\parallel})} = \left(n + \alpha \right) \bar E(p_{\parallel})\,. \label{shift:energy:spectrum}\eeq
 
From (\ref{dirac:quant}) we see that the effect of an external magnetic field of flux $\phi = \alpha' \phi_{0}$ is a further shift of the energy bands, similarly to  twisted PBCs. The combined effect of an external magnetic field and twisted PBCs  (\ref{twisted:PBCs}), is a double energy spectrum shift
\beq
 E_{n}(\bar  p_{\parallel}) = (n + \alpha + \alpha') \bar E (\bar  p_{\parallel})= (n + \alpha + \alpha') \frac {h} {T(\bar  p_{\parallel})}\,.
 \eeq 

This modification of the energy bands associated to an external magnetic field and the resulting modification of the density of states directly affects the properties of the superconducting state. We will study this behavior in the specific case of CNs. 

\subsection{{Coherent state to normal state transition}}\label{coherence}

{  The remaining crucial element to have a consistent description of SC in terms of intrinsic periodicity is the transition between coherent state and normal state, as well as the role of the temperature}. 
 As well-known, for temperatures $\mathcal T$ such that the thermal energy is less than the typical band gap, the material exhibits zero resistivity. Typically, the lower value of band gap scale is fixed by the inverse of the proper time periodicity $T_{\tau} = T(0) $ (representing the extremal value of the time recurrence) whereas the thermal energy is fixed by the inverse of the Euclidean time periodicity $\beta = h / k_{B} \mathcal T$, where $k_{B}$ di the Boltzmann constant \cite{Zinn-Justin:2000dr}. Thus we have that the superconducting phase appears when the characteristic Compton time $T_{\tau} $ is smaller that the thermal periodicity of duration $\beta$ (Euclidean time periodicity).   In fact the quantum  periodicity describes a pure quantum ``coherence'' of the wave-function, whereas the thermal duration $\beta$  describes the dissipation associated to the thermal noise. That is, these two types of temporal periodicities are in competition: the Minkowskian periodicity appears in a real phase of the type $\exp{[- i \bar E (0) T_{\tau} / \hbar]}$ characterizing the recurrence of a pure quantum system (``periodic phenomenon'') whereas the Euclidean periodicity appears in an imaginary phase $\exp{[-  k_{B} \mathcal T \beta / \hbar]}$ describing a dissipation associated to  chaotic interactions \cite{2013arXiv1304.6295F}  (\emph{i.e.} classical or thermal limit). Thus we can also say that if the thermal noise is too big $T_{\tau}/\beta \gg 1$, the  {   collective  ``coherence'' of the quantum periodicity of the ECCs wave-function  (\ref{Period:BCs})} is destroyed by the thermal noise and we have ordinary current. On the other hand, if the quantum recurrence is sufficiently short $T_{\tau}/\beta \ll 1$, the thermal noise cannot destroy the pure ``coherence'' characterizing the quantum recurrence. Only in this pure quantum limit the characteristic aspects of SC described above (\emph{e.g.} flux quantization, zero resistivity, Meissner and Josephon effect, finite step variation of the Goldstone field which transforms as a fermionic pair condensate) becomes manifest. This also implies that high temperature SC occurs in material characterized by short quantum recurrences.  We will use these considerations about the periodicity coherence to interpret the gauge symmetry breaking in SC as an (anomalous) quantum effect induced by BCs. 

\section{Carbon nanotubes}

The above description of SC and related electronic properties in terms of the intrinsic periodicity can be directly tested in CNs, see \cite{RevModPhys.79.677} for a review. 
Experimental evidences and theoretical indications supporting intrinsc SC in CNs are reviewed in \cite{altomare2013one}.  
 According to Bloch's theorem, the properties of the ECCs in the crystalline 2D structure of graphene are determined by the PBCs
  $ \psi(\vec r) \equiv \psi(\vec r + \vec a_1 n_{1} + \vec a_2 n_{1})$,  
   where $\vec a_{1,2} = \{\sqrt 3, \pm 1\} a / 2$, $\{n_{1}, n_{2}\}$ are integer numbers and $a \simeq 2.46\, \AA$, characterizing the electronic properties of the material.  This also implies that the ECCs, at sufficiently low energies, have the dispersion relation typical of a massless particles $\bar E = v_{F} \bar p$, being the speed of light $c$ replaced by the Fermi velocity $v_{F}$ with formation of the typical Dirac cones (for simplicity, in this paper we will not consider the pseudo-spin associated to the honeycomb lattie). 

 Graphene can also be used to form CNs by curling up (compactify) one of the two dimensions of a layer. The geometry of a CN is determined by the compactification vector $\vec C_{h} = n \vec a_{1} + m \vec a_{2} $.  The compactified dimension of the CN constrains the recurrence  of the ECCs along their motion on the graphene lattice, according to the invariant PBCs
 \beq 
 \psi(\vec r) = \psi(\vec r + \vec C_{h})\,. \label{PBCs:graphene}
 \eeq 
 
Besides this PBCs, we must consider that the hexagonal lattice of graphene allows a further intrinsic discrete symmetry for rotations $\frac{2}{3}\pi$. This means that twisted PBCs, analogous to (\ref{twisted:PBCs}), are also admitted
\beq
 \psi(\vec r) = e^{i \frac{ 2 \pi}{3}}\psi(\vec r + \vec C_{h})\,. \label{anti:PBCs:graphene}
 \eeq
 We will refer to (\ref{PBCs:graphene}) as metallic CNs and to (\ref{anti:PBCs:graphene}) as semiconducting CNs. The resulting quantization of magnetic flux is thus given by (\ref{magn:quant}) for $\alpha = 0$ and $\alpha =  \frac 1 3$, respectively.

\subsection{Generation of the effective mass by boundary conditions}

The role of the quantum recurrence of the ECCs  in SC is also clear if we consider that the constraint of periodicity induced by the graphene curled up dimension determine the energy bands structure. 
As already said, the ECCs in infinite graphene layers behave as a massless particles. However, as a dimension is compactified on a circle to form a CN, the ECCs moving in the curled-up direction behave as at rest with respect to the axial direction, \ie they are in the 1D rest frame $p_{\parallel}=0$. That is, even if the ECCs are not moving in the axial directions they have a residual cyclic motion along the radial direction. This residual proper-time periodicity $T_{\tau} = |C_{h}| / v_{F}$  corresponds to the Compton time of the system. Since in undulatory mechanics the mass is linked through the Planck constant to the Compton time,  this rest periodicity determines  the effective mass of the ECCs in the CN, \cite{deWoul:2012ed,RevModPhys.79.677}. Thus, the mass scale determined by the curled-up dimension of the CN, \emph{i.e.} by the PBCs (\ref{PBCs:graphene}) or (\ref{anti:PBCs:graphene}), is 
\beq \bar m =  \frac{h}{ T_{\tau} v_{F}^{2}}\,. \label{found:mass:NT} \eeq

 As the ECC moves along the axial direction with a finite fundamental momentum $\bar p_{\parallel}$, the cyclic motion in this cylindric geometry combined with the axial motion determines a finite spatial  recurrence $\lambda_{\parallel} (\bar p_{\parallel})$ along the axis (which at $\bar p_{\parallel} = 0$ is otherwise infinite)  and a corresponding modulation of temporal  recurrence $T(\bar p_{\parallel})$ such that $T(0) = T_{\tau}$.  In terms of this temporal periodicity the PBCs (\ref{PBCs:graphene}) or  (\ref{anti:PBCs:graphene}) can be equivalently written in the form (\ref{Period:BCs}) and (\ref{twisted:PBCs})  for the collective wave-function used in the general description of the superconducting state. 
   Indeed, \cite{RevModPhys.79.677}, the characteristic energy $\bar E^{2}(p_{\parallel})$  of the ECCs in CNs has a relativistic dispersion relation given by 
   \beq 
   \bar E^{2}(p_{\parallel}) = {\bar m^{2} v_{F}^{4} + \bar p_{\parallel}^{2} v_{F}^{2} }\label{disp:rel:CN}\,,
   \eeq 
    where, as usual, $\bar E (p_{\parallel}) =  {h}/{  T(\bar p_{\parallel}) }$, and $\bar p_{\parallel} =  {h}/{ \lambda_{\parallel}(\bar p_{\parallel}) }$. Hence we have the following the following geometric dispersion relation for the spacetime recurrences $T_{\tau}^{-2} = T^{-2}(\bar p_{\parallel}) - v^2_{F} \lambda^{-2}_{\parallel}(\bar p_{\parallel})$.

   By following the argument of the previous section, we see that the recurrence of the ECCs in CNs  implies a quantization of the magnetic flux, without or with shift of the related spectrum in metallic or semiconducting case respectively, as in (\ref{magn:quant}), \cite{RevModPhys.79.677}. We can therefore extend our general considerations about SC and quantum recurrence to the case of CNs.

\subsection{Electronic properties}

Neglecting the lattice structure, the resulting energy spectrum associated to the temporal recurrence $T(\bar p_{\parallel})$ for the ECCs in their motion along the cundoctor axis is given by (\ref{shift:energy:spectrum}).  Thus, by considering the modulation of periodicity $T(\bar p_{\parallel})$ with $\bar p_{\parallel}$ described in (\ref{disp:rel:CN}), the allowed energy bands $E_{n}$ and the resulting density of state $\rho(E_{\parallel})$ of the ECCs, in the low energy approximation, are given by
\beqa
E^{2}_{n}(p_{ \parallel}) &=& \bar m ^{2} \left(n + \alpha \right)^{2} v_{F}^{4}+ p_{ \parallel}^{2} v^{2}_{F} \label{quan:energy:spectr}\,,  \\
\rho(E_{\parallel}) &=& \frac{\sqrt{3} a^{2}}{T_{\tau} h v^{2}_{F} } \sum_{n} \frac{|E_{n}(p_{ \parallel})|}{\sqrt{E^{2}_{n}(p_{ \parallel}) - m^{2}_{n} v^{4}_{F}}} \label{density:states} \,,
\eeqa
where we have defined $m_{n} = (n + \alpha) \bar m $, and $E_{\parallel} = v_{F} p_{\parallel}$ is the energy in the axial direction,  \cite{RevModPhys.79.677,WhiteCNs}. The overall factor in the density of state $\rho(E_{\parallel})$ is inversely proportional to the Compton time $T_{\tau}$ of the system. Hence, high temperature SC is favored in materials characterized by short time quantum recurrences. 

The effective energy bands of the CNs can be derived from (\ref{quan:energy:spectr}) by considering that these space-time periodicities are not on continuous dimensions but they are on the honeycomb lattice, \cite{deWoul:2012ed,RevModPhys.79.677}. For instance, by considering a time periodicity $T = h / \bar E$ on a lattice, the Fourier transformation implies the quantization of the energy spectrum $E_n = \bar E \frac{N}{\pi} \sin (n \frac{\pi}{N})$. In this way we also note that CNs simulate a periodic Cogwheel model \cite{hooft:2007determinism,hooft:2014cogwheel}. That is, by considering Bloch's theorem (lattice periodicity), in the case of zigzag CN, the condition  (\ref{anti:PBCs:graphene}) yields \cite{deWoul:2012ed}
\begin{equation}\label{nanotube:strectrum}
 E^2_n(\bar {p}_{\parallel})  =  \alpha \bar m^2 v_{F}^4 \frac{N^2}{\pi^2} \left(1 + 2 \cos \frac{\pi n}{N}\right)^{2} - \alpha \bar {p}^2_{\parallel} v_{F}^2 \frac{N^2}{2 \pi^2} \cos \frac{\pi n}{N}\,.\label{ZZ:CN:spectrum}
 \end{equation}
 Notice that all these periodicities are described by the same quantum number $n$ because they all are projections, through the equations of motion, of the rest periodicity of the CN. The fundamental topology of these spacetime recurrences is indeed $\mathbb S^{1}$ (or $\mathbb S^{1}/ \mathcal Z_{2}$ in the Orbifold case). 

By considering the competition between quantum recurrence and thermal dissipation discussed in sec.(\ref{coherence}), we find a heuristic explanation of the fact that the critical temperature of superconducting CN is higher for small diameter: a smaller CN diameter implies a shorter quantum recurrence $T_{\tau}$ of the system with respect to the thermal dissipation time $\beta$ and in turns a higher energy gap with respect to the thermal noise. In fact, while large diameter CNs have a very low critical temperature $\sim 0.55 $K, the small diameter ($\sim 4 \AA$) CNs have a relatively high critical temperature $\sim 15 $ K, \cite{RevModPhys.79.677}. This aspect can be relevant to understand high temperature SC and to identify new materials. 

  A similar example of effective mass generation is in graphene bi-layer. In this case the role of the recurrence is fixed by the periodic correlation of the ECCs among the two layers which result strongly hybridized. 
The role of the Compton length, \emph{i.e.} the analogous of the compactified dimension of the CN, is played by the distance of the layers and the ECCs spatial dynamics is bi-dimensional \cite{2011arXiv1103.1663Z,2013PhRvL.110n6803P}. This separation is $\sim 10^{-10} m$. If compared with the Compton length of the electron $\sim 10^{-12} m$ we find that the effective mass of the ECC in graphene bilayer is $\bar m \sim 10^{-2} \times m_{e}$ as confirmed by direct observations\footnote{{  For a more accurate evaluation we must consider the reduced speed of light in the material}}. By applying a voltage difference $\Delta V$, in agreement with the derivation of the Josephson effect in sec.(\ref{Aharonov:Bohm}), we have the analogous of the twisted PBCs of an angle $2 \pi \alpha = 2 \pi e \Delta V / \bar E(0) = e  \Delta V T_{\tau} / \hbar$. A related shift of the rest energy bands can be obtained with the opening of a superconducting gap between the conducting and the valence bands. 
 { Another example is given by metallic stripes in which the electrons behave as in  potential wells.  In a single metallic stripe, the analogous of the compactified dimension of the CN, \emph{i.e.} the characteristic Compton length, is played by the stripe transversal size determining the effective mass. For electrons at rest with respect to the longitudinal direction of the stripe, the gap is determined by the usual Compton relation. In a superlattice of stripes we have a further Compton scale with additional gaps in the  energy  spectrum determined by the distance between the stripes.}

In a future paper we will extend the phenomenological description of the geometrodynamical  generation of the effective mass in graphene, as well as superfluidity in cigar-shaped ultracold fermions \cite{2012PhRvA..86c3612S} and possible high temperature superconductivity in superlattice of stripes \cite{1996SSCom.100..181P}, in terms of intrinsic periodicity and the related formalism proposed in \cite{Dolce:cycles,Dolce:2009ce,Dolce:tune,Dolce:AdSCFT} and in the present work. 

\subsection{Analogies with the Kaluza-Klein theory}
 
The dispersion relation in graphene single-layers is massless, however the compactification of a dimension to form a CN determines an energy band structure, characterized by a rest energy spectrum $E_{n} (0)$, \ie a mass spectrum\footnote{In particular lattice configurations of Armchair CNs \cite{deWoul:2012ed}, the local extremal point of the dispersion relation determining the Compton time and the mass spectrum of the ECC may occur with small residual the axial momentum.} 
\beq m_{n} = \frac{E_{n}(0)}{v_{F}^{2}} = (n + \alpha) \frac{h}{T_{\tau} v_{F}^{2}} = (n + \alpha) \bar m \,.\label{mass:spectrum}
\eeq 
These are the possible harmonic eigenvalues associated to the twisted Compton (proper-time) periodicity $T_{\tau}$; $\bar m $, defined in (\ref{found:mass:NT}), is the fundamental gap of the mass tower. This is the exact analogous (dual) of the mass spectrum of a classical Kaluza-Klein (KK) theory \cite{Kaluza:1921tu,Klein:1926tv}, as correctly  noted in  \cite{deWoul:2012ed}. In this analogy the  compactification length of the CN $|C_{h}|$ corresponds to the compactification length of the XD. The interpretation of these dualisms is the fact that the rest periodicity, being a recurrence in the worldline of the particle, can be regarded as a cyclic XD with the particularity that, contrarily to the ordinary KK theory, the KK modes are not independent particles but they have a collective behaviors, \ie they are the quantum excitation of the same 4D system, see for instance (\ref{ZZ:CN:spectrum}). In this case we say that the XD is ``virtual'',  \cite{Dolce:AdSCFT}, and the correspondence between CNs and KK theories can be regarded as a generalized aspect of the ``quantum to classical correspondence'' typical of the AdS/CFT correspondence \cite{Witten:1998qj}. As we will discuss in a dedicated paper \cite{Dolce:graphene},  CNs represent the perfect physical system to test the validity at low energies  of field theory in cyclic spacetime dimensions, recently proposed in \cite{Dolce:cycles,Dolce:2009ce,Dolce:tune,Dolce:AdSCFT}.

\section{Considerations about the EM gauge symmetry breaking} \label{Gauge:symmetry:breaking} 

 An interpretation of the gauge symmetry breaking associated to our description of SC stems from the dualism between CNs and XD theories, also pointed out in \cite{deWoul:2012ed}. 
Similarly to XD theories, in CNs the masses are generated by the BCs of a compactified dimension; different BCs correspond to different effective symmetry breaking pattens \cite{Sundrum:2005jf,Csaki:2005vy}.  

As well-known,  in SC the photons exhibit a mass $\bar m^{\gamma}$. This equivalently means that the EM radiation in the superconducting material is characterized by a finite Compton time $T^{\gamma}_{\tau} < \infty$, determined by the mass of the photons  $\bar m^{\gamma} = h / T_{\tau}^{\gamma} c^{2}$. By comparing the Lagrangian of a single massive photon with (\ref{Meissner:expansion})  and (\ref{Meissner:eq})  we find from the quadratic term that the effective Compton length of the photon is of the order of the penetration length $\Lambda \sim h / \bar m^{\gamma} c$. Therefore  the Compton time associated to the  photon mass determines the penetration decay time scale of the magnetic fied in the superconductor.  

The energy quantization of the EM field can be described in terms of the temporal recurrence $T^{\gamma}(\bar  p^{\gamma}_{\parallel})$ of the photon (rather than of the frequency) in a generic reference frame (assuming a continuous lattice with negligible interaction) as $E^{\gamma}_{n}(\bar  p^{\gamma}) =  n \bar E^{\gamma}(\bar  p^{\gamma}_{\parallel}) = n h / T^{\gamma}(\bar  p^{\gamma}_{\parallel})$ --- \emph{e.g.} this is  Planck's quantization of the EM field in the black-body radiation. In other words, we want to interpret the quantized levels of the energy spectrum of a bosonic  system  as the possible vibrational modes associated to the constraint of the corresponding intrinsic periodicity  $T^{\gamma}(\bar  p^{\gamma}_{\parallel})$. The EM field can be therefore expanded in energy eigenmodes allowed by the constraint of intrinsic recurrence
 \beq 
 A_{ \mu} (x)=  \sum_{n=0}^{\infty} A_{n \mu}(\mathbf x, t)  =  \sum_{n=0}^{\infty} A_{n \mu}(\mathbf x) e^{-\frac{i}{\hbar} E^{\gamma}_{n}(\bar { p}_{\parallel}) t}.  \label{EM:expansion}
 \eeq   
 
According to Planck, the energy of the photon is the gap $\bar E( p_{\parallel})$ between the levels of the energy spectrum $E^{\gamma}_{n}(\bar { p}_{\parallel})$ (with possible vacuum energy $\alpha \bar E$ interpretable as a twist in the periodicity).  This also means that, by assuming finite proper-time periodicity $T^{\gamma}(0) = T_{\tau}^{\gamma}$ as for the EM field in SC,  in the rest frame ($ p_{\parallel}=0$) we have a resulting discrete mass (rest energy) spectrum $m^{\gamma}_{n} = n \bar m^{\gamma} = n  h / T_{\tau} c^{2} = E^{\gamma}_{n}(0)/c^{2} = n \bar E^{\gamma}(0)/c^{2}$ associated to the Compton recurrence. Indeed the mass of the photon $\bar m^{\gamma} = h / T_{\tau}^{\gamma} c^{2}$ can be interpreted as the gap between the levels of the mass spectrum $m^{\gamma}_{n}$ associated to  $T^{\gamma}_{\tau}$. \footnote{It is possible to show in several different ways that the PBCs of the matter field induces a periodicity to the related gauge field (modulo gauge transformation), as for instance  by considering the effect of the PBCs of the gauge connection  (\ref{gauge:connection}) or by using the similar formalism of field theory at finite temperature (Euclidean periodicity), as shown in \cite{Zinn-Justin:2000dr}.  A detailed description of this aspect is given in  \cite{Dolce:tune}.}
 
 
Thus, in a semi-classical description analogous to the quantized energy levels of (\ref{EM:expansion}),  the photons of mass $\bar m^{\gamma}$ in superconductors can be regarded as forming a tower $m^{\gamma}_{n} = n \bar m^{\gamma} = n h / T^{\gamma}_{\tau}$ of massive quantum excitations. Their  collective behavior can be therefore described by the following semi-classical action 
  \beq
\mathcal S^{Semi-class.}_{EM} 
\!= \! \sum_{n=0}^{\infty} \!  \int d^{4} x \left [- \frac{1}{4} F_{n \mu\nu} F_{n}^{\mu\nu} - \frac{1}{2} \frac{m^{2}_{n}c^{2}}{\hbar^{2}} A_{n \mu} A_{n}^{\mu} \right]\,. \label{energy:deconstr}
  \eeq 
This semi-classical action is formally equivalent to the classical action of an EM gauge theory defined on a cyclic XD of length $c T_{\tau}$ as can be easily proven by dimensional decompactification, see \cite{Sundrum:2005jf,Csaki:2005vy}. However, in our case, the harmonics describe the collective quantum excitations of the same 4D state whereas in XD theories they describe different classical particles of mass $m_{n}$ (KK modes). That is, in this dualism the cyclic XD must be interpreted as  ``virtual'', \ie it encodes the Compton periodicity of a massive quantum elementary system (\emph{e.g.} the quantum recurrence of an elementary particle) as described in \cite{Dolce:AdSCFT}. 

\subsection{The role of the boundary conditions and outlooks}

Our semi-classical analysis of massive photons suggests that the gauge symmetry breaking occurring in SC can be interpreted as induced by BCs, similarly to XD theories \cite{Sundrum:2005jf,Csaki:2005vy}. The shift of the mass spectrum induced by the twisted-PBCs is the analogous of the Scherk-Schwarz mechanism \cite{Scherk197961}. In terms of the analogies with XD theories, it corresponds to the Hosotani mechanism \cite{Hosotani1983309} at the base of the gauge-Higgs unification, where the twist factor is obtained by allowing a vacuum expectation value to the fifth component of a XD gauge field. 
The role of the vacuum expectation value, which in the Higgs mechanism \cite{PhysRev.117.648,Goldstone1961,PhysRevLett.13.508,PhysRevLett.13.585} is  associated to the mexican hat potential, in our case is played by the fermionic condensation induced, through the quantized Goldstone modes,  by the constraint of intrinsic proper time recurrence. In other words,  { in SC the condensation is strongly characterized by the lattice geometry of the material as well as a possible influences of the intrinsic periodicities of the internal degrees of freedom (\emph{e.g.} the intrinsic recurrences of the electrons in the atomic orbitals)}. 

It is interesting to note that, according to (\ref{eq:periodic:goldstone:phase}) the Goldstone field is periodic (modulo phase factors) so that, for a finite proper time periodicity, it can be expanded in a tower of massive eigenmodes similarly to the matter and gauge fields. A formal analysis of the gauge symmetry breaking, shows that these massive Goldstone  modes provide the extra degrees of freedom  eaten by the  EM gauge field to form gauge eigenmodes of mass $m_{n}$. In fact, these massive eigenmodes of the Goldstone field are formally dual to the discrete massive vibrational eigenmodes of the fifth component of the gauge field in XD theories giving mass to the KK gauge modes, \cite{Sundrum:2005jf,Csaki:2005vy}.

{ This correspondence allows us a further digressions about the phenomenological analogies between the gauge symmetry breaking in SC and that of  some of the most investigate extensions standard model. As well-known from XD gauge theories \cite{Sundrum:2005jf,Csaki:2005vy,Arkani-Hamed:2001ca,Son:2003et}, the action (\ref{energy:deconstr}) (with infinite summation) is gauge invariant despite the mass terms. In fact, by dimensional (de)compactification, , as long as all its  possible mass eigenmodes are considered,  (\ref{energy:deconstr}) is dual to a gauge invariant 5D theory on an interval, which is also 4D gauge invariant, see \cite{Sundrum:2005jf}. 
As in the typical interpretation of the gauge symmetry breaking in XD theories, if the system is observed at low energies , \ie at low temperatures $T^{\gamma}_{\tau} / \beta \ll 1$ such that the thermal energy is less than the band gap scale $\bar m^{\gamma}$, (according to Boltzmann distribution) only the lower quantum mode of the gauge field with non vanishing mass  can be excited (we have already discussed that the in SC massless mode can be neglected  as it corresponds to zero flux). Since only this massive gauge mode is relevant in the effective low  temperatures description of (\ref{energy:deconstr}), we have a theory with explicitly broken gauge symmetry. In agreement with sec.(\ref{coherence}) this also means that at low temperature $T^{\gamma}_{\tau} / \beta \ll 1$ the tower of photons condensate into this massive ground state, and the quantum coherence of the wave-function is not destroyed by the thermal noise. In the opposite limit,  the quantum recurrence $T_{\tau}$ is destroyed by the thermal noise, the corresponding mass spectrum $m_{n}$ cannot form, and the gauge invariance is restored. Thus, we conclude again that high temperature SC corresponds to high energy gap and in turns to materials characterized by short quantum recurrences. 

For a periodicity on a lattice with $N$ sites the action (\ref{energy:deconstr}) turns out to be dual to a moose model with $N$ sites, as can be proven through the dimensional (de)construction mechanism \cite{Arkani-Hamed:2001ca,Son:2003et}, and the related gauge symmetry breaking can be interpreted in analogy with composite-Higgs and technicolor models. }

{ The  collective behavior of the KK modes (``virtual'' KK modes) in (\ref{energy:deconstr}) is typical of holography description of XD theories. As discussed  in \cite{Dolce:AdSCFT},  the dualism between intrinsic periodicity and  XD dynamics allows us to interpret the ``classic to quantum correspondence'' of AdS/CFT and holography in terms quantum recurrence  --- \emph{e.g.} the density of state (\ref{density:states}) is the analogous of the holographic correlator or of the hadronic form-factor in QCD \cite{Son:2003et}. This aspect suggests a possible relationship of our description to holographic SC \cite{Domenech:2010nf}.}

 The interpretation of SC and the corresponding gauge symmetry breaking is fundamental for the understanding of the Higgs mechanism \cite{PhysRev.117.648,Goldstone1961,PhysRevLett.13.508,PhysRevLett.13.585} and the origin of particle masses. The detection and characterization of the Higgs boson is a problem of current interest and fundamental importance not only in high energy physics but also in condensed matter systems such as in superfluids and superconductors, see \cite{PhysRevLett.109.010401} and references therein. Indeed we have seen that a description of particles as elementary cycles suggests us new elements to explore a possible quantum-geometrodynamical origin of particle masses and therefore it can be of interest to investigate with our approach the nature of the Higgs-like boson recently observed at LHC, looking for connections with condensed matter.

\section{Conclusions} 

In this paper we have applied the prescription defined in \cite{Dolce:cycles,Dolce:2009ce,Dolce:tune} (inspired by 't Hooft's Cellular Automata \cite{hooft:2007determinism,hooft:2014cogwheel}) to study fundamental properties of SC, such as magnetic flux quantization, Meissner effect, Josephon effects, the Little-Parks effect,  and energy gap opening.
 Though the microscopic BCS theory is a well established and successful theory, with a very rich phenomenology only partially investigated here, the result of this paper is of heuristic and phenomenological interest because it points out a consistent  macroscopic description of SC based on fundamental aspects of QM, such as  intrinsic recurrence of the pure quantum systems, rather than on the microscopical structure of the material. 
The intrinsic periodicity of the ECCs in a material fully determines its energy band structure (neglecting strong many-body correlations). The energy gap between the bands is fixed, through the Planck constant, by the inverse  of the time recurrence. SC appears when the quantum recurrence is shorter than the time scale associated with the thermal dissipation, $T / \beta \ll 1$. 
Thus this description implies that SC at high critical temperatures is associated with short characteristic time recurrences of the ECCs in the  material. Our approach can be well applied to describe the coherent properties of high temperature superconductors, on the basis of the results here presented. 
 Our description can be directly tested in superconducting CNs  (as well as in graphene  bi-layers  and metallic stripes) in which the intrinsic recurrence turns out to be directly fixed by the cylindric geometry of the material (or by the layers distance or by the transversal size and periodicities of the stripes), determining a finite Compton time and thus the effective mass and the density of the states of the ECCs.  
 
The gauge symmetry breaking occurring in SC is at the base of the Higgs mechanism.   Intrinsic periodicity has allowed us interesting considerations about the role of the BCs, in analogy with gauge theories on a compact XD, and of the quantum dynamics in such gauge symmetry breaking. 


\section*{Acknowledgement} 

 Partial supported from the University of Camerino under
the project FAR ``\emph{Control and enhancement of superconductivity by engineering materials at the nanoscale}''.

\end{document}